\documentclass[aps,prb,reprint,ttisuperscriptaddress]{revtex4}
\usepackage[dvips]{graphicx}
\usepackage{subfigure}
\usepackage{bm}
\usepackage{color}
\usepackage{amsmath,amssymb}
\begin{document}
\title{%Antiferromagnetic spin-wave Doppler shift in strong and weak limits of inter-sublattice electron dynamics
Spin-transfer torques in antiferromagnets: efficiency and quantification method
}
\author{Y. Yamane$^1$, J. Ieda$^{1,2}$, and Jairo Sinova$^{1,3}$}
\affiliation{$^1$Institut f\"{u}r Physik, Johannes Gutenberg Universit\"{a}t Mainz,D-55099 Mainz, Germany}
\affiliation{$^2$Advanced Science Research Center, Japan Atomic Energy Agency, Tokai, Ibaraki 319-1195, Japan}
\affiliation{$^3$Institute of Physics ASCR, v.v.i., Cukrovarnicka 10, 162 53 Praha 6, Czech Republic}
\date{\today}
\begin{abstract}
We formulate a theory of spin-transfer torques in antiferromagnets, which covers the small to large limits of the exchange coupling energy relative to the kinetic energy of the inter-sublattice electron dynamics.
Our theory suggests a natural definition of the efficiency of spin-transfer torques in antiferromagnets in terms of well-defined material parameters, revealing that the charge current couples predominantly to the antiferromagnetic order parameter and the sublattice-canting moment in, respectively, the limits of large and small exchange coupling. 
The effects can be quantified by analyzing the antiferromagnetic spin-wave dispersions in the presence of charge current:
in the limit of large exchange coupling the spin-wave Doppler shift always occurs, whereas, in the opposite limit, the only spin-wave modes to react to the charge current are ones that carry a pronounced sublattice-canting moment.
The findings offer a framework for understanding and designing spin-transfer torques in antiferromagnets belonging to different classes of sublattice structures such as, e.g., bipartite and layered antiferromagnets.
\end{abstract}
\maketitle
%=========================================================================
%           Introduction
%=========================================================================
\section{Introduction}
The conservation of angular momenta between itinerant electrons and localized magnetizations in magnetic materials leads to the fascinating concept of spin-transfer torque (STT)\cite{stt};
the spin angular momentum of the electrons can be transferred to the magnetization via their mutual exchange coupling, which enables to drive the dynamics of magnetization by charge current.
The STT in ferromagnets (FMs), providing a vital information-writing technology, has been driving the explosive growth of the field of spintronics up until now.\cite{stt_review}
In textured FMs, the efficiency of the STT (in the unit of velocity) can be defined by
\begin{equation}
{\bm u}  =  \frac{ g \mu_B P}{ 2 e M_{\rm S} } {\bm j}_{\rm c}  , 
\label{u} \end{equation}
with $g$ the g factor, $\mu_B$ the Bohr magneton, $e$ the elementary charge, $M_{\rm S}$ the saturation magnetization, ${\bm j}_{\rm c}$ the charge current density, and $P$ the net spin polarization carried by the charge current.

Recently, antiferromagnets (AFMs) are generating more attention due to their potential to become a key player in technological applications where AFMs play active roles.\cite{afm_review}
If Eq.~(\ref{u}) is directly applied to AFMs, one would conclude that there can be no STT in AFMs where $P$ becomes zero or vanishingly small;
recent research has been confirming that this is of course not the case.
The study of STTs involving AMF materials was started by investigation of current-driven effects in spin valves or multi-layer systems where each AFM layer carries a single domain.\cite{macdonald,urazhdin,helen,linder,saidaoui,cheng_valve,xu}
Theoretical studies have unveiled an important role of the STT also in textured AFMs as in textured FMs\cite{xu,duine,hals,cheng,tveten,barker};
Xu {\it et al.}\cite{xu} examined the current-driven dynamics of a domain wall (DW) in a two-sublattice AFM metal by {\it ab initio} calculations.
Swaving and Duine\cite{duine} formulated a STT in a one-dimensional bipartite AFM, based on the Landau-Lifshitz (LL) equations for the sublattice-magnetizations in the continuous limit.
Hals {\it et al.}\cite{hals} derived the possible forms of STTs that are allowed by symmetry argument.
The dynamics of AFM textures driven by spin-polarized current has been also studied\cite{helen,cheng}

Thus far, however, it still remains an open question how the STT efficiency, the counterpart of Eq.~(\ref{u}), can be defined for general AFM magnetic textures.\cite{duine,barker}
Finding the STT efficiency would guide us to how to control the STTs in AFMs for designing more prominent STT effects.

In this work, we develop a formalism of current-driven dynamics of two-sublattice AFM textures, where the STT efficiency is provided in terms of unambiguous material parameters.
A challenge in deriving the STT in the AFMs comes from the fact that the electron spin dynamics is not as obvious as in FMs at all, because in the AFM there are two exchange fields corresponding to the two sublattice-magnetizations, that the electron spin can respect.
We formulate the STT in two regimes where the analytical expressions for the electron spin are available;
when the inter-sublattice electron dynamics is dominant over the electron-magnetization exchange coupling, and the opposite.
We find that the STT mechanism that governs its efficiency can quite differ in those two regimes.
In the limit of large exchange coupling, the STT can be generated due to spatial variation of the antiferromagnetic order.
In the opposite limit, on the other hand, the STT requires a sufficiently large canting between the sublattice-magnetizations.
These predictions can be quantified by studying the response of AFM spin waves to the charge current.
In the limit of large exchange coupling, the charge current inevitably causes the spin-wave Doppler shift, whereas, in the opposite limit, it can modify the spin-wave spectrum only when there exists a pronounced sublattice-canting.
Our results demonstrate quantitatively that the STT effects in an AFM highly depend on which class of AFM we consider.

%=========================================================================
%
%           Formalism
%
%=========================================================================
\section{Formalism}
%=========================================================================
%           Model
%=========================================================================
\subsection{Model}
We consider an itinerant AFM composed of two sublattices (1 and 2) with equal saturation magnetization $M_{\rm S}$.
In order to treat the magnetization classically, the coarse graining for the magnetic channel is performed.\cite{neel}
The classical vector ${\bm m}_1 ({\bm r},t)$ $(|{\bm m}_1({\bm r},t)|=1)$ is a continuous function in space that represents the local magnetization direction in the sublattice 1, with a similar definition for ${\bm m}_2({\bm r},t)$;
here the lattice structure is smeared out and the magnetizations of both sublattices are defined at every point in space.
This classical treatment is allowed when the spatial variation of each magnetization is sufficiently slow compared to the atomistic length scale.
The dynamics of the magnetizations are assumed to obey the coupled LL equations with the Gilbert-type damping term\cite{gurevich};
\begin{equation}
\partial_t {\bm m}_i  =  - \gamma {\bm m}_i \times {\bm H}_i
                                    + \alpha {\bm m}_i \times \partial_t {\bm m}_i
                                    + \bm{{\cal T}}_i  ,
                                    \quad  \left( i = 1 , 2 \right)  ,
\label{llg} \end{equation}
where $\gamma$ is the gyromagnetic ratio and $\alpha$ is the damping constant, which are assumed for simplicity to be sublattice independent.
${\bm H}_i = - ( 1 / \mu_0 M_{\rm S} ) \delta w / \delta {\bm m}_i$ are the effective magnetic fields with $w$ being the magnetic energy density, and $\bm{{\cal T}}_i$ are the STTs to be determined.

For the conduction electron channel we employ the following four-band Hamiltonian density\cite{yamane};
\begin{eqnarray}
{\cal H}  &=&  \left( \begin{array}{cc}  t_{11} ( {\bm p} )  &  t_{12} ( {\bm p} )  \\  t_{21} ( {\bm p} )  &  t_{22} ( {\bm p} )  \end{array} \right)
                   + \left( \begin{array}{cc}  J {\bm \sigma} \cdot {\bm m}_1 ( {\bm r} , t )  &  0  \\  0  &  J {\bm \sigma} \cdot {\bm m}_2 ( {\bm r} , t )  \end{array} \right)  \nonumber \\
             &=&  \gamma_0 J {\bm \sigma} \cdot {\bm n} + \left( t_{11} + J {\bm \sigma} \cdot {\bm m} \right) + \gamma_5 t_{12}  ,
\label{h} \end{eqnarray}
a derivation of which starting from an atomistic tight-binding model is discussed in Appendix A.
The upper-left (bottom-right) bands correspond to the sublattice 1 (2).
In the first equality of Eq.~(\ref{h}), the first matrix is the kinetic energy tensor where the diagonal and off-diagonal components describe the intra- and inter-sublattice electron dynamics, respectively, with ${\bm p}$ being the momentum operator of the electron, whereas the second matrix represents the exchange interaction with $J$ being the exchange coupling energy and ${\bm \sigma}$ the Pauli matrices indicating the electron spin operator.
In the second equality, we set $t_{11} = t_{22}$ and $t_{12} = t_{21}$ reflecting the sublattice symmetry, use the tensor product representation of the sublattice and spin spaces with the Dirac matrices
\begin{equation}
\gamma_0 = \sigma_z \otimes I  ,  \qquad
\gamma_5 = \sigma_x \otimes I  , 
\end{equation}
and define the net moment and the N\'{e}el-order vector by
\begin{equation}
{\bm m} = \frac{ {\bm m}_1 + {\bm m}_2 }{ 2 } , \qquad
{\bm n} = \frac{ {\bm m}_1 - {\bm m}_2 }{ 2 } .
\end{equation}
The AFM coupling between ${\bm m}_1$ and ${\bm m}_2$ is the leading energy scale so that $| {\bm m} | \ll 1$ and $| {\bm n} | \simeq 1$.

We regard $J$ and $\langle t_{12} \rangle$ as parameters, where $\langle ... \rangle$ denotes the expectation value at the Fermi surface.
The expressions for $\bm{{\cal T}}_i$ are to be derived in the two limiting cases;
the parameter regimes where $\langle t_{12} \rangle / J \ll 1$ (the exchange-dominant regime hereafter) and where $\langle t_{12} \rangle / J \gg 1$ (the mixing-dominant regime hereafter).
The explicit forms of $t_{11}$ and $t_{12}$ can be determined based on an atomistic tight-binding model, as discussed in Appendix A.
%=========================================================================
%           Exchange-dominant regime
%=========================================================================
\subsection{Exchange-dominant regime}
The condition $\langle t_{12} \rangle/J \ll1$ can be met in AFMs where the inter-sublattice electron dynamics is relatively unfavorable;
e.g., layered AFMs with the c axis being longer than the other axes (Fig.~2b in Appendix A).
To expand ${\cal H}$ in powers of $J^{-1}$, we perform the unitary transformation\cite{yamane}
\begin{equation}
{\cal H}_J  \equiv  e^{ S_J } \left( {\cal H} + i \hbar \partial_t \right) e^{ - S_J }  ,
\label{hj} \end{equation}
with
\begin{equation}
S_J  =  \frac{ t_{12} {\bm \sigma} \cdot {\bm n} }{ 2 J } \gamma_0 \gamma_5  .
\label{sj} \end{equation}
Because the kinetic energy operators, $t_{11}$ and $t_{12}$, in general do not commute with ${\bm m}$ and ${\bm n}$, there appear in Eq.~(\ref{hj}) terms that contain their commutators.
These terms and the last term in Eq.~(\ref{hj}) can be ignored when the spatiotemporal variations of the magnetizations are sufficiently slow (see Appendix B for quantitatively more accurate discussion).
With this condition the expression for ${\cal H}_J$ can be reduced to 
\begin{equation}
{\cal H}_J  =  \left( \begin{array}{cc}  t_{11} + J {\bm \sigma} \cdot {\bm m}_1  &  0  \\  0  &  t_{11} + J {\bm \sigma} \cdot {\bm m}_2 \end{array} \right) + {\cal O} \left( J^{ - 2 } \right)  .
\label{hj2} \end{equation}
Eq.~(\ref{hj2}) proves that the inter-sublattice band-mixing can be neglected up to the order of $J^{-1}$ in the certain condition.
In this rotated frame, the conduction electrons only couple to either ${\bm m}_1$ or ${\bm m}_2$, whereas it is important to note that these sublattice moments are mutually coupled.
Therefore, the spin gauge fields for the itinerant electrons that reside in the $i$-th sublattice are determined by ${\bm m}_i$ and the STTs $\bm{{\cal T}}_i$ in Eq.~(\ref{llg}) are derived  as
\begin{equation}
\bm{{\cal T}}_i  =  \left( {\bm u}_J \cdot \nabla \right) {\bm m}_i 
                          - \beta_J {\bm m}_i \times \left( {\bm u}_J \cdot \nabla \right) {\bm m}_i  ,
\label{stt_j} \end{equation}
where $\beta_J$ is a dimensionless parameter\cite{zhang-li} and the STT efficiency ${\bm u}_J$ is given by
\begin{equation}
{\bm u}_J  =  \frac{ g \mu_B P_{\rm sub} }{ 2 e M_{\rm S} } {\bm j}_{\rm c}  , 
\label{uj} \end{equation}
with $P_{\rm sub}$ representing the spin polarization of the conduction electrons in each sublattice.
We  remark here that Eqs.~(\ref{hj2}) and (\ref{stt_j}) cannot be obtained just by assuming the condition $\langle t_{12} \rangle / J \ll 1$;
if the magnetizations change their directions in time and space fast enough, it can cause considerable inter-sublattice band mixing even in the exchange-dominant regime (see Appendix B).
But still, one should point out that this formal result does justify ignoring the interband hopping and translating things as the STT in each sublattice to be fairly independent. 
%=========================================================================
%           Mixing-dominant regime
%=========================================================================
\subsection{Mixing-dominant regime}
The inter-sublattice electron dynamics may be predominant as $\langle t_{12} \rangle / J \gg 1$ in, e.g., bipartite AFMs where the nearest-neighbor atomic sites connect the different sublattices (Fig.~2a in Appendix A).
We show here that the expressions for STTs in this parameter regime quite differ from Eq.~(\ref{stt_j}).

Let us first perform the following unitary transformation on the sublattice space of Eq.~(\ref{h});
\begin{equation}
{\cal H}_t  \equiv  U {\cal H} U  =  \gamma_0 t_{12} + \left( t_{11} + J {\bm \sigma} \cdot {\bm m} \right) + \gamma_5 J {\bm \sigma} \cdot {\bm n} ,
\end{equation}
with
\begin{equation}
U = ( \sigma_x \otimes I + \sigma_z \otimes I) / \sqrt{2} .
\end{equation}
In the new framework $t_{12}$ comes in the diagonal components, while $J {\bm \sigma} \cdot {\bm n}$ is in the off-diagonal components.
The upper-left (bottom-right) part of ${\cal H}_t$ corresponds to the anti-bonding (bonding) electron states formed by the two sublattice-states.

Then we perform another unitary transformation to expand ${\cal H}_t$ in powers of the operator $t_{12}^{-1}$;
\begin{equation}
{\cal H}_t'  \equiv  e^{ S_t } \left( {\cal H}_t + i \hbar \partial_t \right) e^{ - S_t }  ,
\label{ht2}\end{equation}
with
\begin{equation}
S_t  =  \frac{ t_{12}^{-1} J ( {\bm \sigma} \cdot {\bm n} ) }{ 2 } \gamma_0 \gamma_5  .
\label{st} \end{equation}
Assuming the sufficiently slow and smooth variation in the directions of magnetizations, we can express ${\cal H}_t'$ as (see Appendix C for quantitatively more accurate discussion)
\begin{equation}
{\cal H}_t'  =  \gamma_0 t_{12} + \left( t_{11} + J {\bm \sigma} \cdot {\bm m} \right) + {\cal O} \left( t_{12}^{-2} \right)  .
\label{ht3} \end{equation}
Here we have succeeded in block-diagonalizing ${\cal H}_t$ up to the order of $t_{12}^{-1}$.
In the mixing-dominant regime with the Hamiltonian~(\ref{ht3}), the conduction electron spins only see the net moment ${\bm m}$ regardless of the sublattice degree of freedom.  
In AFMs, ${\bm m}$ can emerge due to several origins such as external magnetic fields, the Dzyaloshinsky-Moriya interaction (DMI), and the spatiotemporal variations in the magnetizations.\cite{andreev,papa}

The magnitude of the net moment $|{\bm m}|$ ($\ll1$) generally varies in both time and space.
This fact makes it difficult to obtain analytical expressions for the STTs for general cases.
In the perfect compensation, i.e., when $| {\bm m} | \rightarrow 0$, the electron-magnetization interaction in Eq.~(\ref{ht3}) vanishes and no STTs arise.
When $| {\bm m} |$ becomes as large as $J | {\bm m} | / \hbar \gg | \partial_t ( {\bm m} / | {\bm m} | ) |$ and $J | {\bm m} | / \hbar \gg | {\bm v}_F \cdot \nabla ( {\bm m} / | {\bm m} | ) |$ over the relevant sample region, it can induce the net spin polarization where the majority (minority) electron spins adiabatically follow the direction of $- {\bm m}$ ($+ {\bm m}$).
In this latter case, $\bm{{\cal T}}_i$ in Eq.~(\ref{llg}) are given by (see Appendix D for a derivation)
\begin{equation}
\bm{{\cal T}}_i  =  - {\bm m}_i \times \left[   \hat{{\bm m}} \times \left( {\bm u}_t \cdot \nabla \right) \hat{{\bm m}}
                                                                + \beta_t \left( {\bm u}_t \cdot \nabla \right) \hat{{\bm m}}
                                                         \right]  , 
\label{stt_t} \end{equation}
where $\hat{{\bm m}} = {\bm m} / | {\bm m} |$, $\beta_t$ is a phenomenological parameter, and the STT efficiency ${\bm u}_t$ is defined by
\begin{equation}
{\bm u}_t  =  \frac{ g \mu_B P_m }{ 2 e M_{\rm S} | {\bm m} | } {\bm j}_{\rm c}  .
\label{ut} \end{equation}
Here, $P_m$ is the net spin polarization of the conduction electrons with respect to ${\bm m}$.
Notice that Eq.~(\ref{stt_t}) clearly differs from Eq.~(\ref{stt_j});
both ${\bm m}_1$ and ${\bm m}_2$ enter the STTs $\bm{{\cal T}}_i$ in Eq.~(\ref{stt_t}), in the contrast that each of ${\bm m}_i$ appears in Eq.~(\ref{stt_j}).
%=========================================================================
%
%           Spin-wave Doppler shift
%
%=========================================================================
\section{Spin-wave Doppler shift}
Here let us study the effects of the STTs on the spin-wave dispersions of the two systems shown in Fig.~1;
an easy-axis (EA) AFM with external dc magnetic field applied along the easy axis, and an easy-plane (EP) AFM with external dc field applied in the easy-plane.
We take the magnetic energy density as\cite{bogdanov}
\begin{equation}
w = A_0 {\bm m}_1 \cdot {\bm m}_2 + A_1 \sum_{ \mu = x , y , z } \left[ ( \partial_\mu {\bm m}_1 )^2 + ( \partial_\mu {\bm m}_2 )^2 - 2 \partial_\mu {\bm m}_1 \cdot \partial_\mu {\bm m}_2 \right] - K ( m_{1z}^2 + m_{2z}^2 ) + \mu_0 {\bm H} \cdot ( {\bm m}_1 + {\bm m}_2 ),
\label{w} \end{equation}
 where $A_0$ and $A_1$ characterize the homogeneous and inhomogeneous exchange couplings, $K$ is the uniaxial anisotropy constant along the $z$ axis, and ${\bm H}$ is the external magnetic field.

In the case of EA-AFM ($K > 0$), both ${\bm m}_1$ and ${\bm m}_2$ lie in the $z$ direction at equilibrium (Fig.~1a) when the external dc field ${\bm H}_{\rm dc}\parallel\hat{{\bm z}}$ is in the range of $0<\omega_H<\sqrt{(2\omega_E+\omega_K)\omega_K}$\cite{gurevich}, where
\begin{equation}
\omega_H \equiv \gamma | {\bm H}_{\rm dc} |, \qquad
\omega_E \equiv \frac{ \gamma A_0 }{ \mu_0 M_{\rm S} } ,  \qquad
\omega_K \equiv \frac{ 2 \gamma K }{ \mu_0 M_{\rm S} } .
\end{equation}
In the absence of charge current, the low-energy spin-wave dispersions of this EA-AFM are given by
\begin{equation}
\omega^{\rm EA}_{{\bm q},\pm}  =  \sqrt{ ( \Lambda {\bm q}^2 + \omega_K ) ( 2 \omega_E + \omega_K ) } \pm \omega_H,
\end{equation}
where
\begin{equation}
\Lambda \equiv \frac{ 4 \gamma A_1 }{ \mu_0 M_{\rm S} } .
\end{equation}
For the EP-AFM ($K < 0$), the parallel component of the magnetizations with respect to the dc field is determined by\cite{gurevich} (Fig.~1c)
\begin{equation}
\sin\varphi_p = \frac{ \omega_H }{ 2 \omega_E } .
\end{equation}
The low-energy spin-wave dispersions are
\begin{eqnarray}
\omega^{\rm EP}_{{\bm q},1}  &=&  \sqrt{  \omega_H^2 \{ 1+ ( | \omega_K | / 2 \omega_E ) \}
                                                                + \Lambda {\bm q}^2 ( 2 \omega_E + | \omega_K | ) \cos^2\varphi_p
                                                      } , \\
\omega^{\rm EP}_{{\bm q},2}  &=&  \sqrt{  2 \omega_E | \omega_K | \cos^2\varphi_p
                                                    + ( \Lambda {\bm q}^2 \sin\varphi_p )^2
                                                    + \Lambda {\bm q}^2 ( 2 \omega_E \cos^2\varphi_p + | \omega_K | \sin^2\varphi_p )
                                                      }.
\end{eqnarray}

\begin{figure}
\centering
\includegraphics[width=8cm,bb=0 0 884 760]{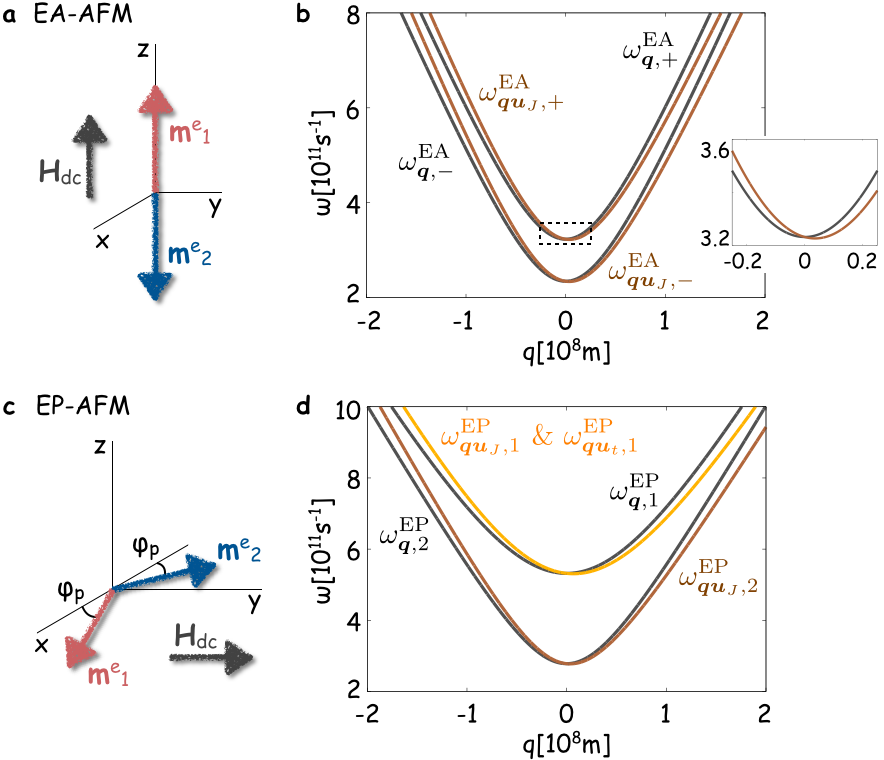}
\caption{ {\bf a}. Schematic of the easy-axis (EA) AMF.
{\bf b}. The spin-wave dispersions of the EA-AFM in the absence ($\omega^{\rm EA}_{ {\bm q}, \pm }$) and presence ($\omega^{\rm EA}_{ {\bm q} {\bm u}_J, \pm }$) of charge current.
The horizontal axis indicates the parallel component of ${\bm q}$ with respect to the charge current.
These modes are affected by the charge current only in the exchange-dominant regime.
The inset magnifies the area indicated by the dotted box, clearly showing the shift of the spectrum around the ${\bm q}=0$ point.
{\bf c}. Schematic of the easy-plane (EP) AMF.
{\bf d}. The spin wave dispersions of the EP-AFM in the absence ($\omega^{\rm EP}_{ {\bm q}, 1 }$ and $\omega^{\rm EP}_{ {\bm q}, 2 }$) and presence ($\omega^{\rm EP}_{ {\bm q} {\bm u}_J, 1 }$, $\omega^{\rm EP}_{ {\bm q} {\bm u}_J , 2}$, and $\omega^{\rm EP}_{ {\bm q} {\bm u}_t, 1 }$) of charge current.
In the mixing-dominant regime, only the $\omega^{\rm EP}_{{\bm q},1}$-mode couples to the charge current.
In {\bf a} and {\bf c}, the arrows ${\bm m}_1^{\rm e}$ and ${\bm m}_2^{\rm e}$ indicate the equilibrium configurations of ${\bm m}_1$ and ${\bm m}_2$.
}
\label{fig01}
\end{figure}

Let us examine the STT effects on the above eigenfrequencies.
In the exchange-dominant regime, Eq.~(\ref{stt_j}) indicates that, in the small dissipation limit with $\alpha\rightarrow0$ and $\beta_J\rightarrow0$, applying the charge current is to replace the partial derivative $\partial_t$ in Eq.~(\ref{llg}) by the Lagrange derivative;
\begin{equation}
{\cal D}_t \equiv \partial_t - {\bm u}_J \cdot \nabla ,
\end{equation}
implying the Galilean invariance of the system with respect to the electron flow.
In systems with the Galilean invariance being respected, the spin-wave spectrum exhibits the current-induced Doppler shift\cite{duine,doppler};
the spin-wave dispersions change as
\begin{equation}
\begin{split}
& \omega^{\rm EA}_{ {\bm q} {\bm u}_J , \pm}  \equiv \omega^{\rm EA}_{{\bm q},\pm} + {\bm u}_J \cdot {\bm q}  , \\ 
& \omega^{\rm EP}_{ {\bm q} {\bm u}_J , 1(2)}  \equiv  \omega^{\rm EP}_{{\bm q},1(2)} + {\bm u}_J \cdot {\bm q}  ,
\end{split}
\quad \left( \langle t_{12} \rangle / J \ll 1 \right) .
\label{eigen_ea-j} \end{equation}

In the mixing-dominant regime, the analytic form for the STTs in Eq.~(\ref{stt_t}) requires the net moment ${\bm m}$ to be large enough to satisfy the condition discussed in the previous section.
For the $\omega^{\rm EA}_{ {\bm q}, \pm }$-modes of the EA-AFM, ${\bm m}$ is mostly vanishingly small and thus tangible STT effects cannot be expected.
For the EP-AFM, on the other hand, there exists the canting moment ${\bm m}$ that can satisfies the above-mentioned condition when the dc field is sufficiently large.
It is shown from Eqs.~(\ref{llg}) and (\ref{stt_t}) that the spin-wave Doppler shift takes place in the $\omega^{\rm EP}_{ {\bm q}, 1 }$-mode but not in the $\omega^{\rm EP}_{ {\bm q}, 2 }$-mode;
\begin{equation}
\begin{split}
&\omega^{\rm EP}_{ {\bm q} {\bm u}_t, 1 }  \equiv  \omega^{\rm EP}_{ {\bm q}, 1 } + {\bm u}_t \cdot {\bm q} ,  \\
&\omega^{\rm EP}_{ {\bm q} {\bm u}_t, 2 }  \equiv  \omega^{\rm EP}_{ {\bm q}, 2 }  ,
\end{split}
\quad \left( \langle t_{12} \rangle / J \gg 1 \right) .
\end{equation}
This distinct feature arises because in the mixing-dominant regime the charge current couples only to ${\bm m}$;
the excitation in the $\omega^{\rm EP}_{{\bm q},2}$-mode is the precession of $(n_x, n_z)$, whereas it is the precession of $(m_x , m_z)$ in the $\omega^{\rm EP}_{{\bm q},1}$-mode.

In Fig.~1b and d compared are the spin-wave dispersions of the EA- and EP-AFMs with and without charge current.
For the material parameters, values in the typical range for AFMs are employed\cite{gurevich}: $A_0 = 2 \times 10^7$ J/m$^3$, $A_1 = 3 \times 10^{-12}$ J/m, $K = 2 \times 10^4$ J/m$^3$, $M_{\rm S} = 8 \times 10^5$ A/m, and $\gamma = 2.215 \times 10^5$ s$^{-1}$/(A/m).
The magnitude of dc field is set to $| {\bm H}_{\rm dc} | = 2 \times 10^5$ A/m and $2.4\times10^6$ A/m for the EA- and EP-AFMs, respectively.
For the ratio of the STT efficiencies, $| {\bm u}_J | / | {\bm u}_t | = 1$ is assumed for simplicity, with $| {\bm u}_J | = | {\bm u}_t | = 300$ m/s.

These results demonstrate the important role played by the inter-sublattice electron dynamics;
the reaction of an AFM to the charge current qualitatively differs depending on the ratio $\langle t_{12} \rangle / J$.
The spin-wave Doppler shift offers a way to quantify the STT in the stationary condition in both time and space.\cite{doppler}
%=========================================================================
%
%           Discussions
%
%=========================================================================
\section{Discussions and Conclusions}
Let us compare our results with existing literature.
For this purpose, we rewrite Eq.~(\ref{llg}) in terms of $({\bm m},{\bm n})$.
In the exchange-dominant regime, this leads to the closed equation of motion for ${\bm n}$;
\begin{eqnarray}
{\bm n} &\times& \left[ ( {\cal D}^2_t - \Lambda \omega_E \nabla^2)  {\bm n} + \gamma^2 ( {\bm n} \cdot {\bm H}_0 ) {\bm H}_0
+ \gamma {\bm n} \times {\cal D}_t {\bm H}_0 
- 2 \gamma ({\bm n} \cdot {\bm H}_0 ) {\bm n} \times {\cal D}_t {\bm n}
 \right. \nonumber \\ && \left.
 -2  \omega_E \omega_K n_z \hat{{\bm z}}
 + 2 \omega_E ( \alpha \partial_t - \beta_J {\bm u}_J \cdot\nabla ) {\bm n}  \right]= 0  ,
\label{n}\end{eqnarray}
while ${\bm m}$ is determined as a slave function of ${\bm n}$;
\begin{equation}
{\bm m} = - \frac{1}{2\omega_E} {\bm n} \times \left(  {\cal D}_t{\bm n}
+ \gamma  {\bm n} \times {\bm H}_0 \right)  ,
\label{m} \end{equation}
where the condition $| {\bm m} | \ll 1$ has been used.
The charge current enters Eqs.~(\ref{n}) and (\ref{m}) through the Lagrange derivative ${\cal D}_t$ (except for the dissipation part), being consistent with the previous discussion regarding the Galilean invariance.
In the absence of charge current, Eqs.~(\ref{n}) and (\ref{m}) reproduce the well-known equations of motion for ${\bm n}$ and ${\bm m}$ under magnetic fields.\cite{andreev}
In the mixing-dominant regime, the $({\bm m},{\bm n})$-representation of Eq.~(\ref{llg}) is generally not as compact as Eqs.~(\ref{n}) and (\ref{m}).
When we limit ourselves to the special case where ${\bm m}$ and ${\bm n}$ are always in a single plane, however, the Galilean invariance is restored in the strict manner, and Eqs.~(\ref{n}) and (\ref{m}) hold with ${\bm u}_J$ and $\beta_J$ replaced by ${\bm u}_t$ and $\beta_t$, respectively.
This condition can be met when, e.g., a DW is formed in a nanowire that possesses a homogeneous DMI with its DMI vector pointing out-of-plane.\cite{bary}
The DW motion predicted by Eq.~(\ref{n}) is consistent with the results in literature (see Appendix E).\cite{duine,hals,tveten}

Eq.~(\ref{n}) contains the STT terms predicted in Ref.~[\onlinecite{hals}] by symmetry argument, whereas the phenomenologically introduced coefficients are now explicitly given by the STT efficiency ${\bm u}_J$.
Since the newly-added terms in Eq.~(\ref{n}) are higher order in terms of the field and derivatives, they were discarded in the previous work.
Eq.~(\ref{n}) not only makes clear that there is the Galilean-invariant nature in the AFMs, but also predicts the cross terms of magnetic field and charge current, which we will investigate elsewhere.

The main focus of Ref.~[\onlinecite{duine}] is on the one-dimensional bipartite AFM where $\langle t_{12} \rangle / J \gg 1$.
They conjectured the nonequilibrium electron spin density proportional to ${\bm n} \times ( {\bm v} \cdot \nabla ) {\bm n}$ with ${\bm v}$ being a parameter in the unit of velocity.
We found that, however, the electron spins predominantly couple to ${\bm m}$ in this parameter regime.
This fact leads to the difference in the results obtained by the two approaches.
While our STT in the mixing-dominant regime is of the first order of $P_m \propto J$, their STT in Ref.~[\onlinecite{duine}] is of higher order as $\propto J^3$.

%=========================================================================
%           Conclusions
%=========================================================================
In conclusion, we have derived the STT efficiency in the two-sublattice AFMs when the inter-sublattice kinetic energy of the conduction electrons is dominant/negligible compared to the exchange coupling energy.
In reality, many of AFM materials should be somewhere in between the two extremes, where numerical approaches will become more powerful.
Our theory demonstrates quantitatively that the STTs in AFMs can, in contrast to in FMs, highly depend on the nature of kinetic energy of the electrons.
These predictions may be tested by studying the spin-wave Doppler shift in the presence of charge current.

%=========================================================================
%           Acknowledgments
%=========================================================================
The authors are grateful to K. Yamamoto, K. Kubo, and M. Mori, for fruitful discussions.
This research was supported by Research Fellowship for Young Scientists from Japan Society for the Promotion of Science, Grant-in-Aid for Scientific Research (No.~24740247, 26247063, 16K05424) from MEXT, Japan, Alexander von Humboldt Foundation, the Ministry of Education of the Czech Republic (Grant No. LM2011026) and the Grant Agency of the Czech Republic (Grant No. 14-37427).
%%%%%%%%%%%%%%%%%%%%%%%%%%%%%%%%%%%%%%%%%%%%%%%%%
%%%%%%%%%%%%%%%%%%%%%%%%%%%%%%%%%%%%%%%%%%%%%%%%%
%%%%%%%%%%%%%%%%%%%%%%%%%%%%%%%%%%%%%%%%%%%%%%%%%
%%%%%%%%%%%%%%%%%%%%%%%%%%%%%%%%%%%%%%%%%%%%%%%%%
%%%%%%%%%%%%%%%%%%%%%%%%%%%%%%%%%%%%%%%%%%%%%%%%%
%%%%%%%%%%%%%%%%%%%%%%%%%%%%%%%%%%%%%%%%%%%%%%%%%
%%%%%%%%%%%%%%%%%%%%%%%%%%%%%%%%%%%%%%%%%%%%%%%%%
%%%%%%%%%%%%%%%%%%%%%%%%%%%%%%%%%%%%%%%%%%%%%%%%%
%%%%%%%%%%%%%%%%%%%%%%%%%%%%%%%%%%%%%%%%%%%%%%%%%
%%%%%%%%%%%%%%%%%%%%%%%%%%%%%%%%%%%%%%%%%%%%%%%%%
%%%%%%%%%%%%%%%%%%%%%%%%%%%%%%%%%%%%%%%%%%%%%%%%%
%=========================================================================
%
%           Appendices
%
%=========================================================================
\section{Appendices}
%=========================================================================
%           A
%=========================================================================
\subsection{Derivation of Eq.~(\ref{h}) from a tight-binding model}
\begin{figure}
\centering
\includegraphics[width=8cm,bb=0 0 865 768]{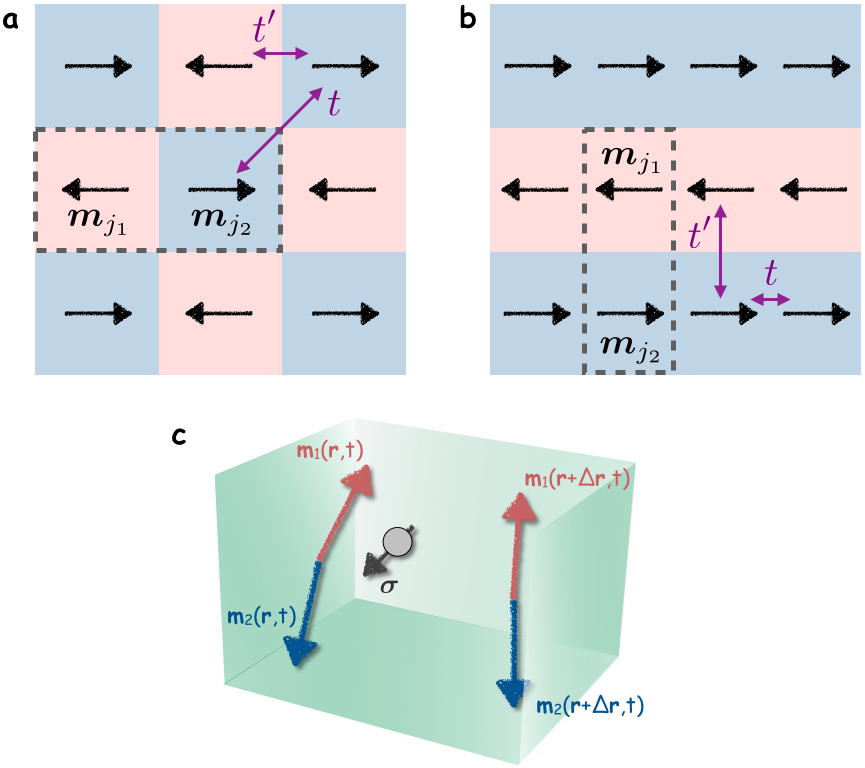}
\caption{ 
{\bf a} and {\bf b}. Schematics of the AFMs with bipartite and layered sublattice structures, respectively.
The dotted boxes indicate the $j$-th unit cells, where the sublattice 1 (2) contributes the magnetization ${\bm m}_{j_1}$ (${\bm m}_{j_2}$).
$t$ and $t'$ represent the nearest-neighbor hopping parameters between intra- and inter-sublattice sites, respectively.
{\bf c}. Schematic of the coarse-grained model in Eq.~(\ref{h}), where both of the sublattice-magnetizations ${\bm m}_1$ and ${\bm m}_2$ are continuous and defined at every point in space.
Such as the atomistic lattice structures and the electron hopping natures are reflected in the kinetic energy tensor of the electron.
 }
\label{fig02}
\end{figure}
Here let us start from an atomistic model for the AFM metal.
The presence of two sublattices leads to unit cells that contain two sites;
the $j$-th unit cell consists of the $j_1$ site that belongs to the first sublattice and the $j_2$ site from the second sublattice, on which the magnetizations ${\bm m}_{j_1}$ and ${\bm m}_{j_2}$ are located, respectively (Fig.~2a and b).
The tight-binding Hamiltonian for the conduction electron is given by
\begin{equation}
{\rm H}  =  t \sideset{}{'}\sum_{ \langle j j' \rangle \sigma } ( c^\dagger_{j_1 \sigma} c_{j'_1 \sigma} + c^\dagger_{j_2 \sigma} c_{j'_2 \sigma} ) 
              + t' \sum_{ \langle j j' \rangle \sigma } ( c^\dagger_{j_1 \sigma} c_{j'_2 \sigma} + {\rm c.c.} )
              + J \sum_{ j \sigma \sigma' } (  c^\dagger_{j_1 \sigma} {\bm \sigma}_{\sigma \sigma'} c_{j_1 \sigma'} \cdot {\bm m}_{j_1}
                                                            + c^\dagger_{j_2 \sigma} {\bm \sigma}_{\sigma \sigma'} c_{j_2 \sigma'} \cdot {\bm m}_{j_2}
                                                            )  . 
\label{h_tight} \end{equation}
Here, $c_{j_1\sigma}$ ($c^\dagger_{j_1 \sigma}$) is the annihilation (creation) operator of an electron with spin $\sigma = \uparrow \downarrow$ at the $j_1$ site, and similarly for $c_{j_2 \sigma}$ ($c^\dagger_{j_2 \sigma}$).
The first and second terms are the kinetic energies, where $t$ ($t'$) represents the nearest-neighbor hopping parameter between intra-(inter-)sublattice sites (Fig.~2a and b).
The sum $\sideset{}{'}\sum$ in the first terms only takes into account the nearest-neighbor intra-sublattice pairs, i.e., $j\neq j'$, whereas the second terms include pairs within unit cells, i.e., $j=j'$.
The third terms describe the on-site exchange coupling.

Introducing the four-component field operator $\Psi_j = ( c_{j_1\uparrow} , c_{j_1\downarrow} , c_{j_2\uparrow} , c_{j_2\downarrow} )^{\rm T}$ and its Fourier transformation by $\Psi_{\bm p} = ( c_{1{\bm p}\uparrow} ,   c_{1{\bm p}\downarrow} , c_{2{\bm p}\uparrow} ,  c_{2{\bm p}\downarrow}   )^{\rm T} \equiv V^{-1/2} \sum_j \Psi_j e^{-i{\bm p}\cdot{\bm r}_j/\hbar}$, with $V$ being the sample volume and ${\bm r}_j$ indicating the position vector of the $j$-th unit cell, Eq.~(\ref{h_tight}) can be rewritten into $4\times4$ fashion as
\begin{equation}
{\rm H}  =  \sum_{{\bm p}} \Psi^\dagger_{\bm p} \left( \begin{array}{cc}  t_{\bm p}  &  t'_{\bm p}  \\  t'_{\bm p}  &  t_{\bm p}  \end{array} \right) \Psi_{\bm p}
              + J \sum_j \Psi^\dagger_j \left( \begin{array}{cc}  {\bm \sigma} \cdot {\bm m}_{j_1}  &  0  \\  0  &  {\bm \sigma} \cdot {\bm m}_{j_2}  \end{array} \right) \Psi_j  .
\label{h_tight2} \end{equation}
Here, $t_{\bm p} \equiv t \sum_{\bm \delta} e^{i{\bm p}\cdot  {\bm \delta}/\hbar  }$ and $t'_{\bm p} \equiv t' \sum_{{\bm \delta}'} e^{i{\bm p}\cdot {\bm \delta}' /\hbar }$, where ${\bm \delta}$ and ${\bm \delta}'$ denote, respectively, the vectors connecting the intra- and inter-sublattice nearest-neighbor sites.
The explicit forms of $t_{\bm p}$ and $t'_{\bm p}$ are given, e.g., in the bipartite AFM by
\begin{equation}
t_{\bm p} = - 4 t a^2 {\bm p}^2 / \hbar^2 , \qquad
t'_{\bm p} = -t' a^2 {\bm p}^2 / \hbar^2 ,
\label{t_bip} \end{equation} 
and in the layered AMF by
\begin{equation}
t_{\bm p} = -t b^2 ( p_x^2 + p_y^2) / \hbar^2 , \qquad
t'_{\bm p} = -t' c^2 p_z^2 / \hbar^2 .
\label{t_lay} \end{equation}
Here $a$ is the lattice constant in the bipartite AFM, and $b$ and $c$ are, respectively, the lattice constants within and between the FM-ordered layers in the layered AFM.
The ${\bm p}$-independent terms have been neglected in Eqs.~(\ref{t_bip}) and (\ref{t_lay}).

By taking the continuous limit in the real space for the second term of Eq.~(\ref{h_tight2}) and moving to the first-quantized representation, we arrive at Eq.~(\ref{h}) where $t_{11}({\bm p})$ and $t_{12}({\bm p})$ are identified with $t_{\bm p}$ and $t'_{\bm p}$, respectively, with ${\bm p}$ read as the quantum operator.
%=========================================================================
%           B
%=========================================================================
\subsection{Derivation of Eq.~(\ref{hj2})}
To expand ${\cal H}$ in Eq.~(\ref{h}) in powers of $J^{-1}$, we perform the unitary transformation in Eq.~(\ref{hj}) or
\begin{eqnarray}
{\cal H}_J  &\equiv&  e^{ S_J } \left( {\cal H} + i \hbar \partial_t \right) e^{ -S_J } \nonumber\\
                         &=&  {\cal H} + [ S_J , {\cal H} ] + \frac{[ S_J, [ S_J, {\cal H} ] ]}{2} + ... + i \hbar \partial_t S_J + ...  , 
\label{hj_ap} \end{eqnarray}
with $S_J$ given in Eq.~(\ref{sj}).
The terms in Eq.~(\ref{hj_ap}) are computed up to the first order of $J^{-1}$ as
\begin{eqnarray}
[ S_J, {\cal H} ]  &=&  - \gamma_5 \left( t_{12} + \frac{ [ t_{12} , {\bm \sigma} \cdot {\bm n} ] {\bm \sigma} \cdot {\bm n} }{ 2 } \right) 
                                  + \frac{ \gamma_0 }{ J } \left( t_{12}^2 {\bm \sigma} \cdot {\bm n} - \frac{ t_{12} [ t_{12} , {\bm \sigma} \cdot {\bm n} ] }{ 2 } \right)  \nonumber \\ &&
                                  + \gamma_0 \gamma_5 \left( - \frac{ t_{12} [ t_{11} , {\bm \sigma} \cdot {\bm n} ] }{ 2 J } 
                                                                                 + \frac{ [ t_{12} , {\bm \sigma} \cdot {\bm m} ] {\bm \sigma} \cdot {\bm n} + 2 i t_{12} {\bm \sigma} \cdot ( {\bm n} \times {\bm m} ) }{ 2 }
                                                                         \right)  ,
\end{eqnarray}
\begin{eqnarray}
\frac{ [ S_J, [ S_J, {\cal H} ] ] }{ 2 }  &=&  - \frac{ \gamma_0 }{ 2 J } \left(   t_{12}^2 {\bm \sigma} \cdot {\bm n} - \frac{ t_{12} [ t_{12} , {\bm \sigma} \cdot {\bm n} ] }{ 2 }
                                                                                                                + \frac{ \left\{ t_{12} {\bm \sigma} \cdot {\bm n} , [ t_{12} , {\bm \sigma} \cdot {\bm n} ] {\bm \sigma} \cdot {\bm n} \right\} }{ 4 } \right)  \nonumber \\ &&
                                                                 - \frac{ {\bm 1} }{ 2 J } \left(   t_{12}^2 {\bm \sigma}\cdot{\bm m} 
                                                                                                           + \frac{ \left[ t_{12} {\bm \sigma}\cdot{\bm n} ,  [ t_{12} , {\bm \sigma}\cdot{\bm m} ] {\bm \sigma}\cdot{\bm n} \right] }{ 4 } 
                                                                                                           + i t_{12} \frac{  [ t_{12} , {\bm \sigma} \cdot ( {\bm n} \times {\bm m} ) ] {\bm \sigma} \cdot {\bm n}
                                                                                                            - [ t_{12} , {\bm \sigma} \cdot {\bm n} ] {\bm \sigma} \cdot ( {\bm n} \times {\bm m} ) }{ 2 } 
                                                                                                    \right)  \nonumber \\ &&
                                                                 + {\cal O} \left( J^{ - 2 } \right)  .
\end{eqnarray}
The expression for ${\cal H}_J$ is thus given by
\begin{equation}
{\cal H}_J  =  \gamma_0 \left( J_{\bm n} {\bm \sigma} \cdot {\bm n} + {\cal F}_1 \right)
                       + {\bm 1} \left( t_{11} + J_{\bm m} {\bm \sigma} \cdot {\bm m} + {\cal F}_2 \right) 
                       + \gamma_5 {\cal F}_3
                       + \gamma_0 \gamma_5 {\cal F}_4
                       + {\cal O} \left( J^{ - 2 } \right)  ,
\label{hj2_ap} \end{equation}
with
\begin{eqnarray}
J_{\bm n}  &\equiv&  J \left( 1 + \frac{ t_{12}^2 }{ 2 J^2 } \right)  , \qquad
J_{\bm m}     \equiv  J \left( 1 - \frac{ t_{12}^2 }{ 2 J^2 } \right)  , 
\label{afm_jn-jm} \\
{\cal F}_1  &=&  - \frac{1}{ 4 J } \left(  t_{12} \left[ t_{12} , {\bm \sigma} \cdot {\bm n} \right]  
                                                        + \frac{ \left\{ t_{12} {\bm \sigma} \cdot {\bm n} , \left[ t_{12} , {\bm \sigma} \cdot {\bm n} \right] {\bm \sigma} \cdot {\bm n} \right\} }{ 2 } 
                                                  \right)  , \\
{\cal F}_2  &=&  - \frac{ 1 }{ 4 J } \left(   \frac{ \left[ t_{12} {\bm \sigma} \cdot {\bm n} , \left[ t_{12} , {\bm \sigma} \cdot {\bm m} \right] {\bm \sigma} \cdot {\bm n} \right] }{ 2 }
                                                           + i t_{12} \left\{  \left[ t_{12} , {\bm \sigma} \cdot \left( {\bm n} \times {\bm m} \right) \right] {\bm \sigma} \cdot {\bm n}
                                                                                   - \left[ t_{12} , {\bm \sigma} \cdot {\bm n} \right] {\bm \sigma} \cdot \left( {\bm n} \times {\bm m} \right)
                                                                                \right\}
                                                    \right)  , \\
{\cal F}_3  &=&  \frac{ \left[ t_{12} , {\bm \sigma} \cdot {\bm n} \right] {\bm \sigma} \cdot {\bm n} }{ 2 }  , \\
{\cal F}_4  &=&  \frac{ t_{12} }{ 2 J } \left( - \left[ t_{11} , {\bm \sigma} \cdot {\bm n} \right]  
                                                                       +  i {\bm \sigma} \cdot \hbar \partial_t {\bm n}
                                                         \right)
                       + \frac{ \left[ t_{12} , {\bm \sigma} \cdot {\bm m} \right] {\bm \sigma} \cdot {\bm n} + 2 i t_{12} {\bm \sigma} \cdot \left( {\bm n} \times {\bm m} \right) }{ 2 }  .
\end{eqnarray}
${\cal F}s$ in Eq.~(\ref{hj2_ap}) can be neglected up to the first order of $J^{-1}$ when the spatiotemporal variations of the magnetizations are as slow as $ \langle \left[ t_{12} , \sigma \cdot {\bm n} \right] \rangle \ll \langle t_{12} \rangle^2 / J$, $ \langle \left[ t_{12} , \sigma \cdot {\bm m} \right] \rangle \ll \langle t_{12} \rangle^2 / J$, $\langle [ t_{11} , {\bm \sigma} \cdot {\bm n} ] \rangle \ll \langle t_{12} \rangle$, and $ \hbar | \partial_t {\bm n} | \ll \langle t_{12} \rangle$.
Approximating both $J_{\bm n}$ and $J_{\bm m}$ by $J$, we arrive at Eq.~(\ref{hj2}).

The commutators of the kinetic energy terms, $t_{11}$ and $t_{12}$, and ${\bm n}$ and/or ${\bm m}$ give rise to a spatial derivative of ${\bm n}$ and/or ${\bm m}$, because the kinetic energies are functions of ${\bm p}=-i\hbar\nabla$.
The expressions of the commutators are accessible by assuming the forms of $t_{11} = {\bm p}^2 / 2 m_{11}$ and $t_{12} = {\bm p}^2 / 2 m_{12}$, where the effective masses $m_{11}$ and $m_{12}$ can be deduced from, e.g., Eqs.~(\ref{t_lay}) in the case of the layered AFM.
Setting $\langle t_{12} \rangle / J = 0.1$ and $J=1$ eV, and employing the typical value for the Fermi wave number $k_{\rm F}$ of the conduction electrons in metals as $k_{\rm F} \sim 10^{10}$ m$^{-1}$, the above mentioned conditions are well met with the spatiotemporal variation of the magnetizations considered in Fig.~1. 
%=========================================================================
%           C
%=========================================================================
\subsection{Derivation of Eq.~(\ref{ht3})}
Expand ${\cal H}_t$ in powers of the operator $t_{12}^{-1}$ by the unitary transformation in Eq.~(\ref{ht2}) or
\begin{eqnarray}
{\cal H}_{t'}  &\equiv&  e^{ S_t } \left( {\cal H}_t + i \hbar \partial_t \right) e^{ - S_t }  \nonumber \\
                           &=&  {\cal H}_t + [ S_t , {\cal H}_t ] + \frac{ [ S_t , [ S_t , {\cal H}_t ] ] }{ 2 } + ...
                                 + i \hbar \partial_t S_t + ...  , 
\label{ht'_ap} \end{eqnarray}
with $S_t$ given in Eq.~(\ref{st}).
Each term in Eq.~(\ref{ht'_ap}) is computed up to the first order of $t_{12}^{-1}$ as
\begin{eqnarray}
[ S_t , {\cal H}_t ]  &=&  - \gamma_5 J \left( {\bm \sigma} \cdot {\bm n} + \frac{ [ t^{-1}_{12} , {\bm \sigma} \cdot {\bm n} ] t_{12} }{ 2 } \right) 
                                      + \gamma_0 J^2 \left( t_{12}^{-1} - \frac{ \left[ t_{12}^{-1} , {\bm \sigma} \cdot {\bm n} \right] {\bm \sigma} \cdot {\bm n} }{ 2 } \right)  \nonumber \\ &&
                                      - \gamma_0 \gamma_5 \frac{ J }{ 2 } \left\{   t_{12}^{-1} [ t_{11} , {\bm \sigma} \cdot {\bm n} ] 
                                                                                                         - J [ t_{12}^{-1} , {\bm \sigma} \cdot {\bm m} ] {\bm \sigma} \cdot {\bm n}  
                                                                                                         - 2 i t_{12}^{-1} J {\bm \sigma} \cdot ( {\bm n} \times {\bm m} )
                                                                                                 \right\}  ,
\end{eqnarray}
\begin{eqnarray}
\frac{ [ S_t , [ S_t , {\cal H}_t ] ] }{ 2 }
&=&  - \gamma_0 \frac{ J^2 }{ 2 } \left(  t_{12}^{-1}
                                                           + \frac{ {\bm \sigma} \cdot {\bm n} [ t_{12}^{-1} , {\bm \sigma} \cdot {\bm n} ] }{ 2 } 
                                                           + \frac{ \left\{ t_{12}^{-1} {\bm \sigma} \cdot {\bm n} , \left[ t_{12}^{-1} , {\bm \sigma} \cdot {\bm n} \right] t_{12} \right\} }{ 4 }
                                                     \right)
         + {\cal O}(t_{12}^{-2}).
\end{eqnarray}
The expression for ${\cal H}_{t'}$ is thus given by
\begin{equation}
{\cal H}_{t'}  =  \gamma_0 \left( t'_{12} + {\cal Y}_1 \right)
                    + {\bm 1} \left( t_{11} + J {\bm \sigma} \cdot {\bm m} \right)
                     - \gamma_5 {\cal Y}_2 + \gamma_0 \gamma_5 {\cal Y}_3  ,
\label{ht'2_ap} \end{equation}
where
\begin{eqnarray}
t'_{12}  &\equiv&  t_{12} \left( 1 + \frac{ t_{12}^{-2} J^2 }{ 2 } \right)  , \\
{\cal Y}_1  &=&  - \frac{ J^2 }{ 4 } \left(   \left[ t_{12}^{-1} , {\bm \sigma} \cdot {\bm n} \right] {\bm \sigma} \cdot {\bm n} 
                                                            + \frac{ \left\{ t_{12}^{-1} {\bm \sigma} \cdot {\bm n} , \left[ t_{12}^{-1} , {\bm \sigma} \cdot {\bm n} \right] t_{12} \right\} }{ 2 }
                          \right)
  , \\
{\cal Y}_2  &=&  - \frac{ J \left[ t_{12}^{-1} , {\bm \sigma} \cdot {\bm n} \right] t_{12} }{ 2 }  ,  \\
{\cal Y}_3  &=&  t_{12}^{-1} J \left(  \frac{ - \left[ t_{11} , {\bm \sigma} \cdot {\bm n} \right] + {\bm \sigma} \cdot i \hbar \partial_t {\bm n} }{2}
                                                               + i J {\bm \sigma} \cdot \left( {\bm n} \times {\bm m} \right)
                                                         \right)
                       + \frac{ J^2 \left[ t_{12}^{-1} , {\bm \sigma} \cdot {\bm m} \right] {\bm \sigma} \cdot {\bm n} }{2}  .
\end{eqnarray}
${\cal Y}$s is Eq.~(\ref{ht'2_ap}) can be neglected up to the first order of $t_{12}^{-1}$ when the spatiotemporal variations of the magnetizations are sufficiently slow that $ \langle [ t_{12}^{-1} , {\bm \sigma} \cdot {\bm n} ] t_{12} \rangle \ll t_{12}^{-2} J^2$, $ \langle [ t_{11} , {\bm \sigma} \cdot {\bm n} ] \rangle \ll t_{12}^{-1} J^2$, and $\hbar | \partial_t {\bm n} | \ll t_{12}^{-1} J^2$.
It can be shown by the similar discussion as in the Appendix B and setting $\langle t_{12} \rangle / J = 10$ and $J=0.1$ eV that these conditions are well satisfied in the systems considered in Fig.~1.
Approximating $t'_{12}$ by $t_{12}$, we arrive at Eq.~(\ref{ht3}).
%=========================================================================
%           D
%=========================================================================
\subsection{Derivation of Eq.~(\ref{stt_t})}
Under the adiabatic approximation where the majority (minority) electron spin adiabatically follows the direction of $-{\bm m}$ ($+{\bm m}$), the (normalized) expectation value ${\bm s}_\pm$ of the conduction electron spin can be represented by
\begin{equation}
{\bm s}_\pm \simeq \mp \hat{{\bm m}} + \delta {\bm s}_\pm  ,
\label{e-spin} \end{equation}
where the upper (lower) sign corresponds to the majority (minority) electron, and $\delta {\bm s}_\pm ( | \delta {\bm s}_\pm | \ll 1)$ is the slight deviation from $\mp \hat{{\bm m}}$.
Assume that the electron spin obeys the following continuity equation;
\begin{equation}
\left( {\bm v}_\pm \cdot \nabla \right) {\bm s}_\pm  =  - \frac{ | {\bm m} | }{ \tau_{\rm ex} } {\bm s}_\pm \times \hat{{\bm m}}
                                                                                   - \frac{1}{ \tau_{\rm sf} } \delta {\bm s}_\pm  ,
\label{eom-s} \end{equation}
where ${\bm v}_\pm$ denotes the average electron velocity, $\tau_{\rm ex} = \hbar / 2 J$, and $\tau_{\rm sf}$ is the relaxation time for the electron-spin flip.

By substituting Eq.~(\ref{e-spin}) into (\ref{eom-s}), the expression for $\delta {\bm s}_\pm$ is obtained as
\begin{equation}
\delta{\bm s}_\pm  =  \pm \frac{ \tau_{\rm ex} }{ | {\bm m} | } \left[   \hat{{\bm m}} \times \left( {\bm v}_\pm \cdot \nabla \right) \hat{{\bm m}}
                                                                                                     + \beta_t \left( {\bm v}_\pm \cdot \nabla \right) \hat{{\bm m}}
                                                                                               \right]  ,
\end{equation}
where $\beta_t = \tau_{\rm ex} / \tau_{\rm sf}$.
The torques ${\bm T}_i$ that the electron spins exert on ${\bm m}_i$  are given by
\begin{eqnarray}
{\bm T}_i  &=&  - \gamma {\bm m}_i \times \left[ - \frac{ 1 }{ \mu_0 M_{\rm S} } J \left( n_+ {\bm s}_+  +  n_- {\bm s}_- \right) \right]  \nonumber \\
                &=&  - \frac{ g \mu_B \left( n_+  -  n_- \right) }{ 2 \tau_{\rm ex} M_{\rm S} | {\bm m} | } {\bm m}_i \times {\bm m}_{j\neq i}
                         - {\bm m}_i \times \left[  \hat{{\bm m}} \times \left( {\bm u}_t \cdot \nabla \right) \hat{{\bm m}}
                                                            + \beta_t \left( {\bm u}_t \cdot \nabla \right) \hat{{\bm m}} \right]  , 
\label{stt-t_ap} \end{eqnarray}
where $n_{ + ( - ) }$ is the majority (minority) electron density, and the STT efficiency ${\bm u}_t$ is given in Eq.~(\ref{ut}).
In the second equality of Eq.~(\ref{stt-t_ap}), $\gamma = g \mu_B \mu_0 / \hbar$ and ${\bm j}_{\rm c} = - e ( n_+ {\bm v}_+  +  n_- {\bm v}_- )$ have been used.
Here, the spin polarization $P_m$ of the conduction electrons with respect to ${\bm m}$ is defined by $P_m {\bm j}_{\rm c} = - e ( n_+ {\bm v}_+  -  n_- {\bm v}_- )$.
The first term in the second equality of Eq.~(\ref{stt-t_ap}) contributes the modulation to the AFM coupling between ${\bm m}_1$ and ${\bm m}_2$;
we absorb this first term into the definition of the AFM exchange coupling.
The second terms in the second equality of Eq.~(\ref{stt-t_ap}), which are Eq.~(\ref{stt_t}), are the STTs $\bm{{\cal T}}_i$ that act on textured AFMs.
%=========================================================================
%           E
%=========================================================================
\subsection{Domain wall motion}
Consider a one-dimensional AFM nanowire stretching in the $z$-axis with easy-axis anisotropy ($K>0$) along it.
An equilibrium AFM texture is determined by ${\bm n}$ and ${\bm m}$ at which the magnetic energy density $w$, which is given in Eq.~(\ref{w}), takes an extremal value.
In the absence of external field, a static DW solution satisfying the boundary condition $n_z(\pm\infty)=\mp1$ is
\begin{eqnarray}
\theta &=& 2 \tan^{-1} [e^{(z-q)/\Delta} ] , \\
\varphi &=& 0  ,
\end{eqnarray}
where the polar angles are defined by ${\bm n}=(\sin\theta\cos\varphi,\sin\theta\sin\varphi,\cos\theta)$, $q$ represents the DW center position, and $\Delta=\sqrt{2A_1/K}$.

Apply a dc charge current in the $z$ direction, and examine the current-driven dynamics of the DW by using Eq.~(\ref{n}) for the exchange-dominant regime.
To obtain an analytical solution for the DW dynamics, we make the steady-motion approximation, where the DW maintains the equilibrium profile with $q$ being time dependent;
the DW exhibits a translational motion described by time evolution of the collective coordinate $q$.
By rewriting Eq.~(\ref{n}) into the equation of motion for $q$ by preforming the volume integral of the equation,\cite{tveten} one obtains
\begin{equation}
\left. \frac{dq}{dt} \right|_{t\rightarrow\infty}  = - \frac{\beta_J}{\alpha} u_J
\end{equation}
As the dynamics of AFM textures in general has an inertia,\cite{andreev,bary} the above equation provides with the terminal velocity of the DW.
The dissipative process described by $\beta_J$ is required to drive the DW by the charge current.
As pointed in the Sec.~IV, the same argument applies to the mixing-dominant regime with $\beta_J$ and $u_J$ replaced by $\beta_t$ and $u_t$, respectively, when there exists a sufficiently large ${\bm m}$ over the relevant sample region and ${\bm n}$ and ${\bm m}$ lie in a single plane.
When, in the mixing-dominant regime, the magnitude and direction of ${\bm m}$ have some significant dependence on time and space that does not meet the above-mentioned conditions, it can make it difficult to obtain analytical expressions for the STT effects, which is beyond the scope of the present paper.
%=========================================================================
%           References
%=========================================================================


\begin{thebibliography}{99}
\bibitem{stt}
J. C. Slonczewski, J. Magn. Magn. Mater. {\bf 159}, L1 (1996);
L. Berger, Phys. Rev. B {\bf 54}, 9353 (1996).
\bibitem{stt_review}
D. C. Ralph and M. D. Stiles, J. Magn. Magn. Mater. {\bf 320}, 1190 (2008);
A. Brataas, A. D. Kent, and H. Ohno, Nature Mater. {\bf 11}, 372 (2012).
\bibitem{afm_review}
T. Jungwirth, X. Marti, P. Wadley, and J. Wunderlich, Nat. Nanotechnol. {\bf 11}, 231 (2016).
\bibitem{macdonald}
A. S. N\'u\~nez, R. A. Duine, P. Haney, and A. H. MacDonald, Phys. Rev. B {\bf 73}, 214426 (2006);
R. A. Duine, P. M. Haney, A. S. N\'u\~nez, and A. H. MacDonald, Phys. Rev. B {\bf 75}, 014433 (2007);
Z. Wei, A. Sharma, A. S. N\'u\~nez, P. M. Haney, R. A. Duine, J. Bass, A. H. MacDonald, and M. Tsoi, Phys. Rev. lett. {\bf 98}, 116603 (2007);
P. M. Haney, D. Waldron, R. A. Duine, A. S. N\'u\~nez, H. Guo, and A. H. MacDonald, Phys. Rev. B {\bf 75}, 174428 (2007).
P. M. Haney and A. H. MacDonald, Phys. Rev. Lett. {\bf 100}, 196801 (2008);
\bibitem{urazhdin}
S. Urazhdin and N. Anthony, Phys. Rev. Lett. {\bf 99}, 046602 (2007).
\bibitem{helen}
H. V. Gomonay and V. M. Loktev, Phys. Rev. B {\bf 81}, 144427 (2010);
H. V. Gomonay, R. V. Kunitsyn, and V. M. Loktev, Phys. Rev. B {\bf 85}, 134446 (2012).
\bibitem{linder}
J. Linder, Phys. Rev. B {\bf 84}, 094404 (2011).
\bibitem{saidaoui}
H. Ben Mohamed Saidaoui, A. Manchon, and X. Waintal, Phys. Rev. B {\bf 89}, 174430 (2014).
\bibitem{cheng_valve}
R. Cheng, J. Xiao, Q. Niu, and A. Brataas, Phys. Rev. Lett. {\bf 113}, 057601 (2014);
R. Cheng, M. W. Daniels, J.-G. Zhu, and D. Xiao, Phys. Rev. B {\bf 91}, 064423 (2015).
\bibitem{xu} Y. Xu, S. Wang, and K. Xia, Phys. Rev. Lett. {\bf 100}, 226602 (2008).
\bibitem{duine} A. C. Swaving and R. A. Duine, Phys. Rev. B {\bf 83}, 054428 (2011): J. Phys.: Condens. Matter {\bf 24}, 024223 (2012).
\bibitem{hals} K. M. D. Hals, Y. Tserkovnyak, and A. Brataas, Phys. Rev. Lett. {\bf 106}, 107206 (2011).
\bibitem{cheng} R. Cheng and Q. Niu, Phys. Rev. B {{\bf 86}}, 245118 (2012): {\it ibid}. {\bf 89}, 081105(R), (2014).
\bibitem{tveten} E. G. Tveten, A. Qaiumzadeh, O. A. Tretiakov, and A. Brataas, Phys. Rev. Lett. {\bf 110}, 127208 (2013).
\bibitem{barker} J. Barker and O. A. Tretiakov, Phys. Rev. Lett. {\bf 116}, 147203 (2016).
\bibitem{neel} L. N\'{e}el, Ann. Phys. (Paris), {\bf 3}, No. 2, 137 (1948).
\bibitem{gurevich} A. G. Gurevich and G. A. Melkov, {\it Magnetization Oscillations and Wave}, (CRC Press, 1996).
\bibitem{yamane} Y. Yamane, J, Ieda, and J. Sinova, Phys. Rev. B {\bf 93}, 180408(R) (2016).
\bibitem{zhang-li} S. Zhang and Z. Li, Phys. Rev. Lett. {\bf 93}, 127204 (2004).
\bibitem{andreev} A. F. Andreev and V. I. Marchenko, Phys. Usp. {{\bf 23}}, 21 (1980).
\bibitem{papa} N. Papanicolaou, Phys. Rev. B {\bf 51}, 15062 (1995); {\it ibid.} {{\bf 55}}, 12290 (1997).
\bibitem{bogdanov} A. N. Bogdanov, U. K. R\"ossler, M. Wolf, and K.-H. M\"uller, Phys. Rev. B {\bf 66}, 214410  (2002).
\bibitem{doppler} V. Vlaminck and M. Bailleul, Science {\bf 322}, 410 (2008).
\bibitem{bary} V. G. Bar'yakhtar, B. A. Ivanov, and M. V. Chetkin, Sov. Phys. Usp. {{\bf 28}}, 7 (1985).
\end{thebibliography}
\end{document}